\documentclass{article}

\usepackage{PRIMEarxiv}

\usepackage[utf8]{inputenc} 
\usepackage[T1]{fontenc}    
\usepackage{hyperref}       
\usepackage{url}            
\usepackage{booktabs}       
\usepackage{amsfonts}       
\usepackage{nicefrac}       
\usepackage{microtype}      
\usepackage{lipsum}
\usepackage{fancyhdr}       
\usepackage{graphicx}       
\graphicspath{{media/}}     
\usepackage{amsmath} 

\title{CheapNET: Improving Light-weight speech enhancement network by projected loss function

}

\author{
    Kaijun Tan \thanks{Equal contribution by both authors.} \\
    SEMI \\
    tkj@semi.ac.cn \\
    \And
    BenZhe Dai \footnotemark[1]\\
    SEMI \\
    \texttt{daibenzhe@gmail.com} \\
    \And
    Jiakui Li \\
    Sense Time Research \\
    \texttt{ljiakui@sensetime.com} 
    \And
    Wenyu Mao \thanks{Corresponding author for this article.} \\
    SEMI \\
    \texttt{maowenyu@semi.ac.cn} 
}

\begin{document}
\maketitle
\begin{abstract}
Noise suppression and echo cancellation are critical in speech enhancement and essential for smart devices and real-time communication. Deployed in voice processing front-ends and edge devices, these algorithms must ensure efficient real-time inference with low computational demands. Traditional edge-based noise suppression often uses MSE-based amplitude spectrum mask training, but this approach has limitations. We introduce a novel projection loss function, diverging from MSE, to enhance noise suppression. This method uses projection techniques to isolate key audio components from noise, significantly improving model performance. For echo cancellation, the function enables direct predictions on LAEC pre-processed outputs, substantially enhancing performance. Our noise suppression model achieves near state-of-the-art results with only 3.1M parameters and 0.4GFlops/s computational load. Moreover, our echo cancellation model outperforms replicated industry-leading models, introducing a new perspective in speech enhancement.

\end{abstract}


\section{Introduction}
Noise suppression and echo cancellation are two crucial subtasks in speech enhancement algorithms\cite{survey}. These algorithms are typically employed at the front end of the signal processing pipeline, particularly at the edge, necessitating a balance between performance requirements and computational complexity in most applications. Current research predominantly focuses on achieving a 0.5 real-time factor on laptop devices\cite{dns2023}, a benchmark under which many studies have demonstrated effective results. 

FULLSUBNET\cite{fullsubnet}, one of the several open-source state-of-the-art (SOTA) solutions, partitions audio into different frequency bands, employing a feature fusion module and a video common mask framework for speech enhancement. FULLSUBNET-plus\cite{chen2022fullsubnet+} further enhances this by incorporating mel-spectrum as an input feature and introducing a channel attention module, significantly improving the model's performance. DCUNET\cite{dcunet} introduces a speech enhancement network based on a complex Codec structure, pioneering the use of Complex CNN modules. Building on this, DCCRN\cite{hu2020dccrn} modifies the feature fusion module into a complex network, achieving substantial performance gains and representing the previous generation SOTA solution. Conv Tasnet\cite{conv-tasnet}, an early CNN-based speech enhancement network, was initially proposed for voice separation and later adapted for speech enhancement. DPRNN\cite{dprnn}introduced a dual-path RNN network, strengthening the RNN's modeling capability within the model and resulting in performance gains. DPTNet\cite{dptnet} employs an end-to-end Transformer model for speech enhancement. \cite{baidu-aec} was the SOTA solution in the AEC-challenge 2022\cite{aec2022}, proposed a hybrid time and frequency attention encoder-decoder structure. This architecture achieves good performance in both speech enhancement and echo cancellation tasks with a relatively small parameter count, but its computational complexity is higher due to the use of attention in the time dimension.

In the realm of lightweight speech enhancement models, research is comparatively scarce. The earliest lightweight speech enhancement algorithm, RNN-noise\cite{rnn_noise}, has a computational complexity of just 0.04G flops/S. Nsnet1\cite{nsnet1} builds on this using RNN as the primary framework and demonstrates good performance in speech enhancement tasks. Nsnet2\cite{nsnet2}, an enhancement of Nsnet1, deepens the network structure and employs new training methods, significantly boosting model performance. These models served as the baseline for DNS2020\cite{reddy2020interspeech} and DNS 2022\cite{reddy2021icassp}, respectively. Further advancements include the Swin-Transformer-based speech enhancement network\cite{swing-ft}, which implements a new network architecture and low-complexity, high-performance speech enhancement using the Swin Transformer\cite{liu2021swin}. In the AEC task, Nsnet1's structure and Nsnet2's training methods served as the baseline for the AEC challenge 2022\cite{aec2022}, outperforming nearly half of all submissions that year. The winning solution, CRUSE\cite{cruse}, utilized an amplitude common mask prediction architecture for speech enhancement, a model now widely implemented across Microsoft's product lines.

Inspired by these works, we believe that models based on amplitude spectrum prediction are sufficiently simple, cost-effective, and industry-accepted\cite{nsnet1,aec2022, cruse}. Consequently, we focused our research on speech enhancement algorithms within this framework. We observed that under this framework, the learning objectives of the model were not precise enough, impacting its noise reduction performance. Furthermore, in AEC tasks, while LAEC features are widely introduced, existing methods struggle to accurately estimate masks on LAEC processed results. Typically, LAEC is used as a reference signal rather than as a masking object, which we believe limits the model's echo cancellation performance. To address these issues, we propose a projection loss function in this work, allowing the model to directly predict the mask object's maximal speech component. This provides a more accurate prediction target for noise reduction tasks and enables direct mask prediction on LAEC results for echo cancellation tasks. Our experiments have demonstrated significant improvements in speech performance.

In this paper, we address two pivotal challenges in the realm of speech enhancement: noise suppression and acoustic echo cancellation. These challenges are particularly relevant in the context of smart devices and real-time communication systems, where the ability to process audio signals efficiently and effectively is paramount. Traditional approaches in these domains have primarily leveraged Mean Squared Error (MSE)-based amplitude spectrum mask training. While this methodology has provided a foundation for progress, it often falls short in addressing the intricacies of real-world audio processing, particularly under the constraints of limited computational resources.

Recognizing these limitations, our work introduces innovative methodologies to advance the state-of-the-art in speech enhancement, particularly focusing on the computational efficiency and efficacy of the models used. Our contributions are threefold and represent a significant leap in the field of speech processing:

Projection Loss Function: We propose a novel projection loss function, a departure from traditional MSE-based methods. This function enables models to more accurately predict the most significant vocal component within noisy audio samples. By focusing on the key elements of speech within a noise environment, this loss function offers a more precise and effective approach to speech enhancement.

Application in AEC Tasks: Extending the utility of our projection loss function, we apply it to Acoustic Echo Cancellation (AEC) tasks. This application allows models to directly predict masks based on pre-processed outputs from the Look-Ahead Echo Cancellation (LAEC) method. Such an approach not only enhances echo cancellation capabilities but also streamlines the processing pipeline, resulting in improved overall performance in AEC tasks.

CHEAPNET Methodology: Addressing the need for computationally efficient yet effective models, we introduce CHEAPNET. This method utilizes simple GRU networks to achieve competitive speech quality enhancement in both noise suppression and echo cancellation subtasks. CHEAPNET demonstrates that robust speech enhancement can be achieved without resorting to complex and resource-intensive models, a crucial consideration for deployment in edge devices and real-time systems.

\section{Method}

\subsection{VAD-Projected Loss Function}

In this work, we innovatively enhance speech by predicting amplitude spectrum masks on noisy audio. This process is mathematically expressed as:
\begin{equation}
Y_{mag}(i,j) = X_{mag}(i,j) \times P_{mag}(i,j)
\end{equation}
Here, $Y_{mag}(i,j)$ represents the target audio component at time $i$ and frequency $j$, while $X_{mag}$ and $P_{mag}$ denote the amplitude spectrum of the original noisy audio and the model's prediction of the proportional mask, respectively.

For the task of noise suppression, the original audio signal $X(t)$ is conceptualized as a linear summation of clean speech $C(t)$ and noise $N(t)$:
\begin{equation}
X(t) = C(t) + N(t)
\end{equation}
Similarly, in echo cancellation tasks, the original signal $X(t)$ is a linear combination of speech $C(t)$, noise $N(t)$, and a reference signal $R(t)$:
\begin{equation}
X(t) = C(t) + N(t) + R(t)
\end{equation}
Given the linearity of the Fourier transform, these signals' spectral representations can also be considered linear combinations. Hence, for noise suppression:
\begin{equation}
X_{spec}(f) = C_{spec}(f) + N_{spec}(f)
\end{equation}
and for echo cancellation:
\begin{equation}
X_{spec}(f) = C_{spec}(f) + N_{spec}(f) + R_{spec}(f)
\end{equation}
In the complex domain, which is also linear, the l2 norm for noise suppression satisfies:
\begin{equation}
\|X_{spec}(f)\| \le \|C_{spec}(f) \| + \| N_{spec}(f) \|
\end{equation}
It's feasible to construct $X_{spec}(f)$ and $C_{spec}(f)$ that adhere to the above equations while fulfilling:
\begin{equation}
\|X_{spec}(f)\| < \|C_{spec}(f) \|
\end{equation}
Considering the amplitude spectrum represented as $X_{mag}(f) = \|X_{spec}(f) \|$, and the fact that slicing the audio signal before performing a Discrete Fourier Transform does not affect these conclusions, we confirm that "there exist original audio signals $X$, speech signals $C$, and time-frequency points $i, j$ such that $X_{mag}(i,j) < C_{mag}(i,j)$" is valid in practice and widely applicable.

Previous work has typically relied on optimizing $arg min \{MSE(C_{mag}, Y_{mag})\}$ to learn the proportional mask $P$ in Equation 1. However, as $P$ ranges between 0 and 1, for cases where inequality 7 holds, the learning target $C_{mag}$ becomes unattainable. We believe this gap significantly impedes the model's ability to learn noise patterns. This phenomenon is also present in AEC tasks and is even more pronounced due to the increased acoustic components in the signal model. Another motivation for addressing this issue is our observation that LAEC pre-processing leaves the target speech components nearly intact while incompletely eliminating the reference signal component. Thus, we aim to directly predict masks on the LAEC results in AEC tasks, where the correlation between LAEC and original signals is weakened, making conventional MSE learning less effective.

We address this issue by shifting the learning target to the projection of $C_{mag}$ onto $X_{mag}$, denoted as $C^{\prime}_{mag}$:
\begin{equation}
C^{\prime}_{mag} = C_{mag}\frac{ X_{mag} \cdot C_{mag}  }{ X_{mag}^2}
\end{equation}
The revised learning objective is then:
\begin{equation}
arg min \{MSE(C^{\prime}_{mag}, Y_{mag})\}
\end{equation}

To further enhance model performance, inspired by [source], we introduced a VAD loss on top of the projected MSE. The training system, illustrated below, integrates this additional component:

\begin{equation}
\text{loss} = MSE(C'_{mag}, VAD(Y_{mag}) \times Y_{mag}) + MSE(X_{mag } - C'_{mag}, X_{mag} \times Y_{mag})
\end{equation}

\begin{figure}
    \centering
    \includegraphics[width=0.8\linewidth]{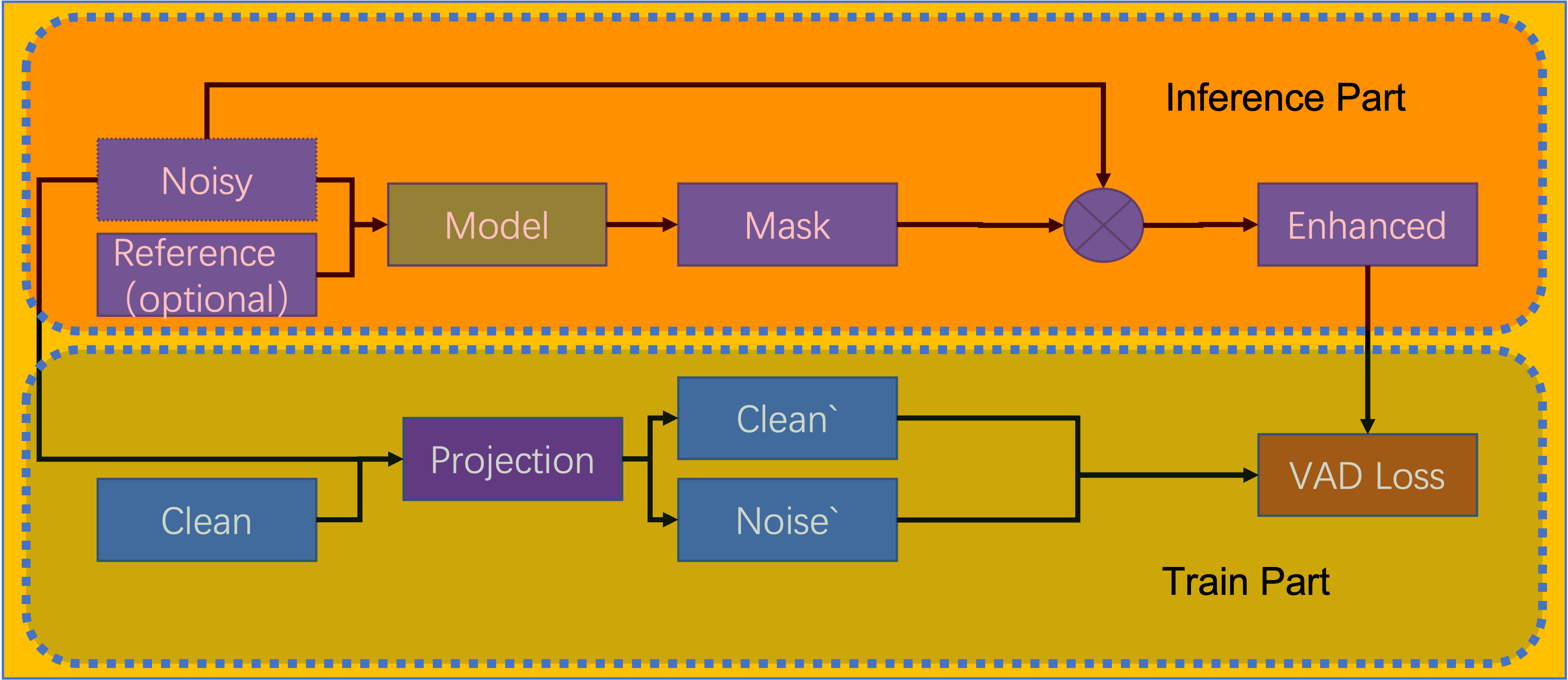}
    \caption{Enter Caption}
    \label{fig:enter-label}
\end{figure}

\subsection{Model Architecture}
In our research, we employ a network architecture centered around a two-layer Gated Recurrent Unit (GRU) with residual connections, serving as the backbone of our model. Preceding the GRU layers, we incorporate a set of linear layers tasked with aligning the input shape to match the GRU cell count. Following the GRU, the architecture includes a Feed-Forward Network (FFN) complemented by an additional linear layer, responsible for mapping the features to the desired output dimensions. At the outermost layer of the model, a sigmoid function is utilized to output percentage values, essential for generating amplitude spectrum masks.

For the Acoustic Echo Cancellation (AEC) task, the input to the model comprises a dual-channel amplitude spectrum, where one channel represents the primary signal, and the other serves as a reference signal. This primary signal may either be a noise-containing audio amplitude spectrum or an amplitude spectrum processed by Linear Acoustic Echo Cancellation (LAEC). The proportionate mask is then applied to the primary signal. In contrast, for the Deep Noise Suppression (DNS) task, the model takes the noise-containing audio as the sole primary signal. The overall structure of our model, designed to effectively handle these tasks, is illustrated below.

\begin{figure}
    \centering
    \includegraphics[width=0.8\linewidth]{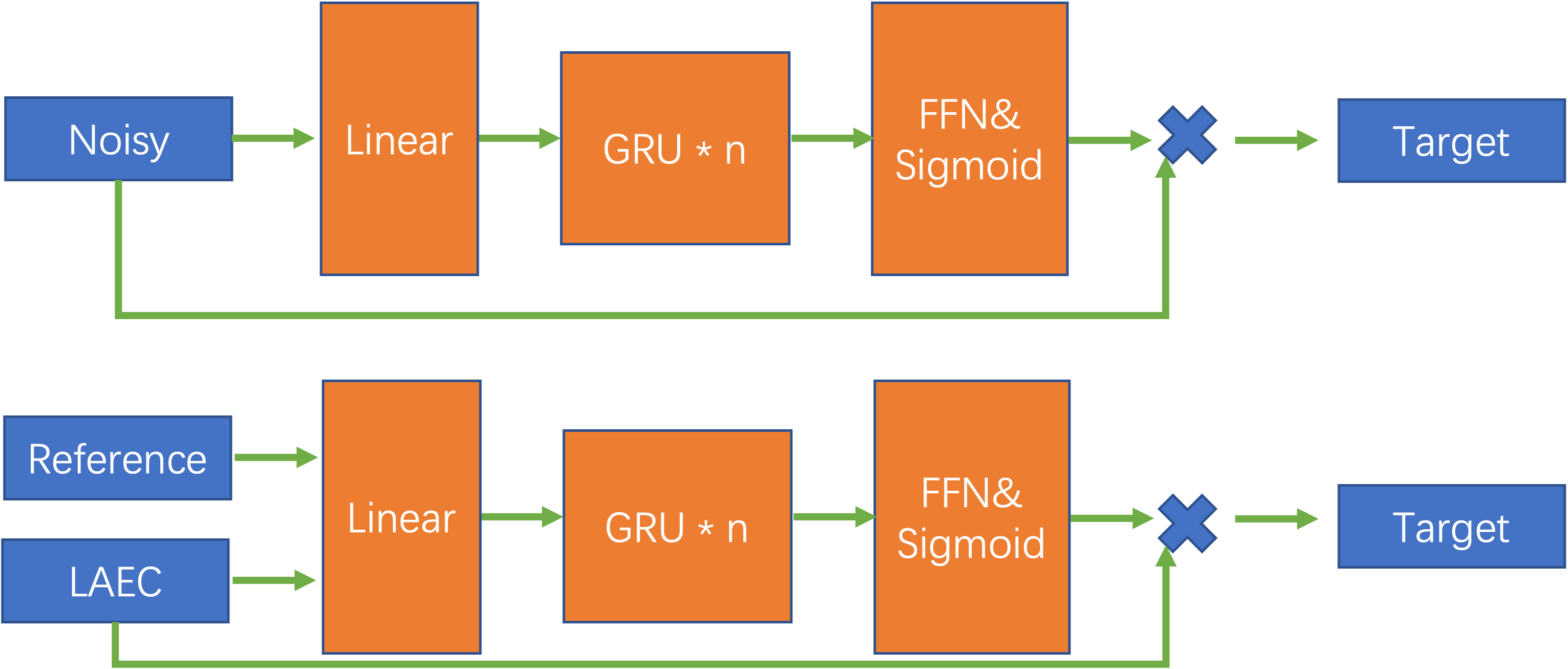}
    \caption{Enter Caption}
    \label{fig:enter-label}
\end{figure}

\section{Experiment}
In this paper, we present our novel models GRU-512 and GRU-256, which we specifically developed for the tasks of Deep Noise Suppression (DNS) and Acoustic Echo Cancellation (AEC). These models stand out for their significantly lower computational complexity – two orders of magnitude less – compared to the state-of-the-art (SOTA) FULLSUBNET model, while maintaining comparable performance. Moreover, our models notably outperform other baseline models, particularly NSNET1, despite having a similar architecture. 

For the DNS task, we utilized the publicly available DNS-challenge\cite{reddy2020interspeech, reddy2021icassp} 16k dataset to compile our training and testing sets. We strictly segregated the voice and noise audio samples into training, validation, and testing sets in an 8:1:1 ratio, ensuring no overlap between them. During training, we adopted a strategy of random sampling and online mixing to ensure data diversity. For testing, we randomly selected voice and noise samples from the test set, mixing them at various signal-to-noise ratios (SNR) to create individual test cases of 10 seconds each. Our training covered SNR ranges from -5 to 15 dB, while the test set included SNR ranges from -15 to 15 dB, with SNRs below 0 dB considered as low SNR and above 0 dB as high SNR.

For the AEC task, we randomly chose two voice audio samples to represent the target and reference signals. These were first mixed at a certain SNR, followed by the addition of a noise signal at a fixed SNR. This process was conducted online during training, while for testing, we used a pre-mixed dataset created based on the same rules. In our study, we assumed a causally related random delay of up to 500ms between the original and reference signals in the AEC task.

In our comparative analysis, we benchmarked our GRU-512 and GRU-256 models against a suite of prominent baseline models known for their proficiency in noise suppression and enhancement. These baseline models, all with accessible open-source weights, might have had a theoretical advantage in our experiments due to possible exposure to our test data during their development. For the DNS task, our selection included NSNET1 and NSNET2, which are recognized as the baselines in the DNS Challenge, as well as DCCRN, DPTNet, DCUNET, and DPRNN, all of which are reproductions from the Asteroid framework based on their original publications.

In our AEC task experiments, the primary baseline model was derived from the AEC Challenge 2023, representing an advanced iteration of NSNET1 tailored for acoustic echo cancellation. This baseline's architecture is meticulously mirrored in our GRU-320 model. The core of our experimental study was centered on the GRU series models, particularly those prefixed with 'LAEC'. These LAEC models are innovatively designed to perform masking predictions on outputs processed through the LAEC method. The primary aim of this experiment was to validate the efficacy of our approach in enhancing results through masking predictions applied to LAEC-processed outputs.

Furthermore, to demonstrate the versatility and applicability of our method across different frameworks, we replicated the CRUSE model, which was Microsoft's submission for the AEC Challenge 2023. This replication served as a crucial aspect of our study, proving that our LAEC approach significantly boosts performance even in models like CRUSE. By doing so, we showcased the broad applicability and effectiveness of our LAEC method in enhancing acoustic echo cancellation capabilities across various model architectures.

Both training and testing datasets were sourced exclusively from the DNS-challenge 16k dataset, with a strict policy of excluding any test audio from the training set. For the AEC task, our training involved a range of Signal-to-Noise Ratio (SNR) and Echo-to-Signal Ratio (ESR) from -5 to 10 dB. In the DNS task, the SNR range was also set from -5 to 10 dB. All training datasets were derived from online mixing. Our experiments did not introduce reverberation in both tasks.

The results of these experiments are detailed in Tables 1 and 2.

\begin{table}[h!]
\centering
\caption{The experiment result on DNS testset}
\begin{tabular}{lllllllll}
    \toprule
Model Name         & STOI Low & STOI High & STOI  & PESQ Low & PESQ High & PESQ  & Parameters & Computation Cost \\ 
    \midrule
noisy              & 0.555        & 0.832         & 0.713        & 1.356        & 1.975         & 1.710        & -                & -                      \\
    \midrule
Fullsubnet-plus\cite{chen2022fullsubnet+}    & 0.661        & 0.897         & 0.796        & 1.735        & \textbf{3.052} & \textbf{2.487} & 8.6M             & 61.2GFlops/s           \\
DCCRN\cite{hu2020dccrn}              & 0.557        & 0.870         & 0.736        & 1.402        & 2.521         & 2.041        & 3.7M             & 13.5GFlops/s           \\
dptnet\cite{dptnet}             & 0.604        & 0.886         & 0.765        & 1.553        & 2.731         & 2.226        & 2.782            & 11.7GFlops/s           \\
dcunet\cite{dcunet}             & 0.622        & 0.883         & 0.771        & 1.497        & 2.681         & 2.173        & 7.65M            & 124.8GFlops/s          \\
dprnn\cite{dprnn}              & 0.603        & 0.887         & 0.765        & 1.539        & 2.716         & 2.212        & 3.63M            & 117.9GFlops/s          \\
nsnet-baseline\cite{nsnet1}     & 0.457        & 0.702         & 0.597        & 1.337        & 2.195         & 1.827        & 3.1M             & 0.2GFlops/s   \\
nsnet2-baseline\cite{nsnet2}    & 0.633        & 0.869         & 0.768        & 1.583        & 2.618         & 2.174        & 6.1M             & 0.4GFlops/s            \\
    \midrule
GRU-512            & \textbf{0.681} & \textbf{0.906} & \textbf{0.810} & \textbf{1.801} & 2.950         & 2.458        & 3.41M            & 0.2GFlops/s            \\
GRU-256            & 0.621        & 0.870         & 0.763        & 1.549        & 2.516         & 2.102        & \textbf{0.92M}   & \textbf{0.06GFlops/s}           \\
    \bottomrule
\end{tabular}
\end{table}

\begin{table}
 \caption{The experiment result on AEC testset}
  \centering
  \begin{tabular}{lcccccc}
    \toprule
    Name     & WB\_PESQ     & NB\_PESQ     & STOI     & ESTOI     & AEC\_MOS     & DEG\_MOS \\
    \midrule
    noisy    & 1.12        & 1.38         & 0.61     & 0.43      & 2.09         & 3.73     \\
    baseline\cite{aec2022,nsnet1}  & 1.24        & 1.59         & 0.69     & 0.52      & 4.07         & 3.03     \\

    \midrule
    LAEC-GRU-320 & 1.50        & 1.96         & 0.79     & 0.64      & 4.16         & 3.07     \\
    GRU-512   & 1.45        & 1.90         & 0.77     & 0.61      & 3.87         & 3.23     \\
    LAEC-GRU-512  & \textbf{1.58}        & \textbf{2.06}         & 0.80     & 0.65      & 4.27         & 3.29     \\
    CRUSE    & 1.30        & 1.70         & 0.73     & 0.54      & 3.65         & 2.83     \\
    LAEC-CRUSE & 1.54        & 2.02         & \textbf{0.80}     & \textbf{0.65}      & \textbf{4.30}         & \textbf{3.29}     \\

    \bottomrule
  \end{tabular}
  \label{tab:table}
\end{table}

\section{Conclusion}
The experimental results for both the DNS\cite{dns2023} and AEC\cite{aec2022,aec2023} tasks in our study offer compelling evidence supporting the effectiveness of our newly proposed projection loss function. This loss function is a pivotal advancement, enabling models to learn more accurate prediction targets.

In the context of the DNS task, our models, GRU-512 and GRU-256, demonstrated remarkable performance. They achieved close proximity to the state-of-the-art FULLSUBNET model in terms of performance, but with significantly lower computational complexity. This is particularly noteworthy as our models managed to substantially outperform other models, especially those with a similar structure like NSNET1. The efficacy of the projection loss function in this task is evident from the improved precision in noise suppression and speech enhancement that our models exhibited.

For the AEC task, the projection loss function showcased additional benefits. It allowed our models to directly predict masks based on LAEC-processed results. The LAEC-prefixed models, such as LAEC-GRU-512, demonstrated this capability effectively. By applying our masking predictions on LAEC-processed outputs, we achieved superior performance in acoustic echo cancellation. Furthermore, the replication of the CRUSE model and its significant performance enhancement with the LAEC method validated the versatility of our approach.

In summary, the experimental outcomes validate our projection loss function's dual advantages: enhancing prediction accuracy in general and facilitating direct mask prediction on LAEC-processed results in AEC tasks. These results underscore the potential of our approach to set new standards in both DNS and AEC tasks, paving the way for more efficient and effective models in the field of audio processing.

\section*{Acknowledgments}
This work is supported by the Chinese Academy of Sciences and the SEMI.
\bibliographystyle{unsrt}  
\bibliography{references}

\end{document}